\documentclass[twocolumn,aps,pra]{revtex4-1}
\usepackage[latin9]{inputenc}
\usepackage{amsmath}
\usepackage{amssymb}
\usepackage{graphicx}

\begin{document}

\title{Two-dimensional Fermi gases near a $p$-wave resonance: \\ effect of quantum fluctuations}

 \author{Shao-Jian Jiang}
 \affiliation{Department of Physics and Astronomy, University of British
 	Columbia, Vancouver V6T 1Z1, Canada}
 \author{Fei Zhou}
 \affiliation{Department of Physics and Astronomy, University of British
 	Columbia, Vancouver V6T 1Z1, Canada}

\date{}
\begin{abstract}
  We study the stability of $p$-wave superfluidity against quantum fluctuations in two-dimensional Fermi gases near a $p$-wave Feshbach resonance .
  An analysis is carried out in the limit when the interchannel coupling is strong.
  By investigating the effective potential for the pairing field via the standard loop expansion, we show that a homogeneous $p$-wave pairing state becomes unstable when two-loop quantum fluctuations are taken into account.
  This is in contrast to the previously predicted $p+ip$ superfluid in the weak-coupling limit [V. Gurarie {\it et al.}, Phys. Rev. Lett. {\bf 94}, 230403 (2005)].
  It implies a possible onset of instability at certain intermediate interchannel coupling strength.
  Alternatively, the instability can also be driven by lowering the particle density.
  We also discuss the validity of our analysis.
\end{abstract}

\maketitle

\section{Introduction}
As the key to electronic superconductivity and fermionic superfluidity, pairing between fermions has been one of the main topics in condensed matter physics.
It can be classified according to the orbital angular momentum of the pair wave function.
Compared to $s$-wave pairing, pairing with finite angular momentum can host richer structures and plays an important role in novel superconductors and superfluids.
For example, $p$-wave pairing in superfluid $^3$He leads to gapless $A$ phase and gapped $B$ phase with different symmetry properties~\cite{Leggett1975}.
Later, a $p$-wave superconductor was also discovered~\cite{Maeno1994}.
Moreover, $p+ip$ pairing in two dimensions (2D) was shown to be topologically nontrivial with excitations exhibiting non-Abelian statistics~\cite{Read2000,Kitaev2001,Alicea2012,Volovik1999,Moore1991,Nayak1996}.
These have been proposed to be a key ingredient for topological quantum computation~\cite{Kitaev2003}.
This has drawn increasing attention to pairing of fermions in the $p$-wave channel.

The ultracold Fermi gas is a promising platform to study $p$-wave pairing.
Although interactions in the $s$-wave channel are usually dominant due to the centrifugal barrier in finite-angular-momentum channels, a Fermi gas interacting in the $p$-wave channel can be realized by preparing fermions in the same pseudo-spin state, where Pauli exclusion suppresses $s$-wave scattering.
Furthermore, the $p$-wave interaction can be enhanced via resonance techniques, e.g., the Feshbach resonance~\cite{Chin2010}.
In laboratories, $p$-wave resonances have been found for $^{40}$K and $^6$Li~\cite{Regal2003,Zhang2004,Schunck2005}. 
The manipulation of interactions in low dimensions has also been investigated~\cite{Guenter2005,Nishida2010,Peng2014}.
These techniques offer amazing tunability of interaction parameters, and makes realizing $p$-wave superfluid in Fermi gases an appealing idea~\cite{Yamaguchi2016,Fedorov2017}.
It is thus necessary to understand $p$-wave Fermi gases from weak- to strong-interaction regime, which in turn can provide inspiring insight into $p$-wave pairing problems in other condensed matter systems.
It has been shown that the $p$-wave superfluid at zero temperature behaves quite differently from the $s$-wave case~\cite{Botelho2005,Ho2005,Gurarie2005,Cheng2005,Iskin2006,Gurarie2007}.
Instead of a crossover, it undergoes a phase transition between different $p$-wave superfluid phases, when the attraction between fermions is adiabatically increased.
The phase diagram can be even richer when anisotropy in the $p$-wave channel is further considered.
Other properties of a $p$-wave Fermi gas including the transition temperature and stability have also been studied~\cite{Ohashi2005,Iskin2006,Gurarie2007,Levinsen2007,Kagan2008,Mizushima2008,Inotani2012}.
Recently, several works have also studied this problem by introducing $p$-wave contacts~\cite{Yoshida2015,Yu2015,He2016,Luciuk2016,Zhang2016a}.
However, in the intermediate-interaction regime near the threshold of a two-body bound state, i.e., resonance, where the length scale associated with the interaction diverges, systematic understanding is still challenging.

A $p$-wave Feshbach resonance is usually narrow~\cite{Ticknor2004,Gubbels2007,Fuchs2008} because the centrifugal barrier generally suppresses the interchannel coupling.
For this reason previous theoretical studies have mainly been focusing on this weak-coupling limit.
In this limit, the fluctuation effect is perturbative in terms of the interchannel coupling even near resonance.
It is thus possible to carry out a controllable analysis with the mean field capturing the leading order contribution, which leads to a gapped $p$-wave superfluid in 2D~\cite{Gurarie2007}.

A recent experimental progress~\cite{Dong2016} shows the possibility of broad $p$-wave resonances in ultracold atomic gases.
Theoretically, however, a systematic understanding of interacting fermions near a strong-coupling $p$-wave resonance is still lacking due to the potential break down of the perturbative approach when the interchannel coupling is strong.
Moreover, a series of cluster states called super-Efimov states have been proposed by Nishida {\it et al.}~\cite{Nishida2013} as a possible candidate for 2D $p$-wave interacting fermions.
The general relation between different correlated states such as the cluster states and superfluid remains to be understood.
In this Letter, we will exclusively discuss the 2D $p$-wave superfluid right at resonance, especially its stability, as a small step towards this direction.

We start with a two-channel Hamiltonian describing a system of spinless fermions interacting in the $p$-wave channel in two dimensions:
\begin{align}
\label{eq:2}
  H ={} & \sum_{\boldsymbol{k}} c_{\boldsymbol{k}}^{\dagger} \left( \frac{k^2}{2} - \mu \right) c_{\boldsymbol{k}} + \sum_{i,\boldsymbol{Q}} b_{i,\boldsymbol{Q}}^{\dagger} \left( \frac{Q^2}{4} - 2 \mu + \delta \right) b_{i,\boldsymbol{Q}}  \nonumber \\
  & + \frac{g}{\sqrt{\Omega}} \sum_{i,\boldsymbol{Q}, \boldsymbol{k}} k_{i} \left( b_{i, \boldsymbol{Q}} c_{\boldsymbol{Q}/2 + \boldsymbol{k}}^{\dagger} c_{\boldsymbol{Q}/2 - \boldsymbol{k}}^{\dagger} + h.c. \right),
\end{align}
where $c_{\boldsymbol{k}},c_{\boldsymbol{k}}^{\dagger}(b_{\boldsymbol{k}},b_{\boldsymbol{k}}^{\dagger})$ are fermionic (bosonic) annihilation and creation operators for fermions in the open channel (molecules in the closed channel), respectively, $\mu$ is the chemical potential, $\delta$ is the energy detuning of the closed channel, $g$ is the interchannel coupling, and $\Omega$ is the volume.
The summation of momentum is up to a ultraviolet (UV) cutoff, $\Lambda$, which parametrizes microscopic details of the system determined by atomic physics.
In our model describing an interacting Fermi gas, $\Lambda$ is the largest momentum scale, i.e., $\Lambda^2 /|\mu| \gg 1$ and $\Lambda / k_F \gg 1$, where $k_F$ is the Fermi wave vector.
We set the reduced Planck constant and the atom mass to be unity throughout the paper.
The bosonic operator is labeled by $i = x,y$, which is related to the annihilation (creation) of $p$-wave molecules with definite angular-momentum projection, $b_{\pm}^{(\dagger)}$, by $b_{\pm}^{\dagger} = \mp (b_x^{\dagger} \pm i b_y^{\dagger})/\sqrt{2}$.

\paragraph{Two-body scattering in the $p$-wave channel.}
Before carrying out a systematic analysis of the stability of a superfluid, we can first get some insight by considering a two-body scattering problem in the $p$-wave channel in 2D.
The corresponding scattering $T$-matrix~\cite{Messiah} can be calculated by considering the Hamiltonian (Eq.~\eqref{eq:2}) in vacuum in 2D:
\begin{equation}
\label{eq:1}
T(E,\boldsymbol{k},\boldsymbol{k}^{\prime}) = \boldsymbol{k} \cdot \boldsymbol{k}^{\prime} \left[ -\frac{\delta}{g^2} + \frac{\Lambda^2}{4 \pi} +  \frac{E}{g^2} + \frac{E}{4 \pi} \ln \frac{\Lambda^2}{E} + i \frac{E}{4} \right]^{-1},
\end{equation}
where $E$ is the scattering energy, $\boldsymbol{k}(\boldsymbol{k}^{\prime})$ is the relative momentum of the initial (final) state.
The quadratic dependence on $\Lambda$ in the denominator of the $T$-matrix can be removed by introducing the scattering area, and resonance occurs when
\begin{equation}
  -\frac{\delta_R}{g^2} = -\frac{\delta}{g^2} + \frac{\Lambda^2}{4\pi} = 0,
\end{equation}
i.e., when the physical or renormalized detuning $\delta_R$ is set to be zero.
Note that the interchannel coupling $g$ can be either strong or weak depending on the specific atoms involved in resonance.
The logarithmic term in Eq.~\eqref{eq:1} , on the other hand, shows that the two-body scattering properties in the $p$-wave channel inevitably depend on short-distance details of the interparticle interaction~\cite{Braaten2012}.
For this reason, an additional contact is needed for the $p$-wave Fermi gas~\cite{Yu2015}.
In this Letter, we will focus on the strong-coupling limit when the interchannel coupling $g$ is large, or more specifically, when
\begin{align}
\label{eq:9}
g^2 \ln \frac{\Lambda}{k_F} \gg 1.
\end{align}
Apparently, this condition can also be satisfied in the low-density limit.
In this limit, the $T$-matrix in Eq.~\eqref{eq:1} for the two-channel model can be reduced to the same structure as the 2D one-channel model if one identifies $(-\delta/g^2+\Lambda^2/(4\pi))^{-1}$ as the scattering area $s$.
One can also confirm that in the same limit, the hybridization between scattering atoms in the open channel and the closed-channel molecules is strong so that the physical molecules have a negligible weight in the closed channel~\cite{weight}.
In order to analyze the problem of superfluidity, we take an approach of the effective potential.
In this approach, the effective potential of the system as a function of $\boldsymbol{\Delta}$, i.e., $E(\boldsymbol{\Delta})$, is first calculated, where $\boldsymbol{\Delta}$ is related to the expectation value of the bosonic field in the closed channel via $\Delta_{i} = g \langle b_{i,0} \rangle / \sqrt{\Omega}$. 
\( \boldsymbol{\Delta} \) is then determined by minimizing $E(\boldsymbol{\Delta})$.
All quantum fluctuations carried by bosonic fields $b^{(\dagger)}_{i, \boldsymbol{Q} \neq 0}$ are encoded in $E$, which can be calculated diagrammatically via quantum loop expansion~\cite{Nishida2006,Jiang2014}.

\section{Mean-field results}
\begin{figure}
  \includegraphics[width=3.5in]{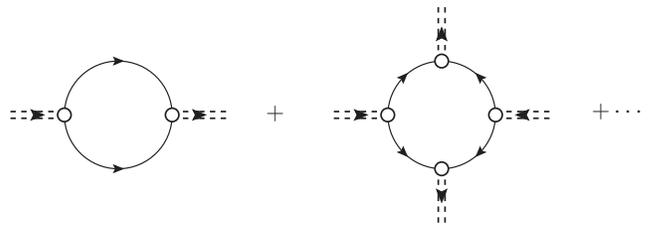}
  \caption{\label{fig:1loop}One-loop diagrams contributing to $E^{(1)}$.
    The solid line and the double dashed line represent the free-fermion propagator and the classical pairing field $\boldsymbol{\Delta}/g$, respectively.
    The open circle is the interchannel coupling vertex, which has the form $g \boldsymbol{k}$ in the $p$-wave channel (see the second line of Eq.~\eqref{eq:2}).}
\end{figure}

First of all, there is a tree-level contribution which can be read out from the Hamiltonian,
\begin{equation}
  \label{eq:tree}
E^{(0)} = g^{-2} (\delta - 2 \mu) |\Delta|^2,
\end{equation}
where $|\Delta|^2 = {\boldsymbol{\Delta}} \cdot \boldsymbol{\Delta}^{*}$.
We then look at the diagrams with one fermionic loop, which do not contain bosonic propagators, as shown in Fig.~\ref{fig:1loop}.
They can be summed up as:
\begin{align}
\label{eq:4}
E^{(1)} & = \frac{1}{2} \int \frac{d^2k}{(2 \pi)^2} \left[  |\xi_{\boldsymbol{k}}| - \sqrt{\xi_{\boldsymbol{k}}^2 + 4 \left|\boldsymbol{\Delta} \cdot \boldsymbol{k} \right|^2}\right],
\end{align}
where $\xi_{\boldsymbol{k}} = \boldsymbol{k}^2/2 - \mu$.
Further adding the energy density of free fermions with chemical potential $\mu$, $E^{\text{free}} =- \mu^2/(4\pi)$, we arrive at the mean-field effective potential, $E^{\text{MF}} = E^{\text{free}} + E^{(0)} + E^{(1)}$.

\begin{figure}
   \centering
   \includegraphics[width=3.5in]{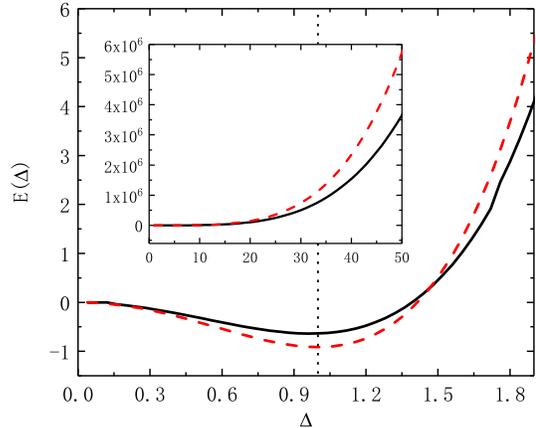}
   \caption{\label{fig:mf}Mean-field effective potential along the direction $\boldsymbol{\Delta} = \Delta/2 (\boldsymbol{e}_x + i \boldsymbol{e}_y)$ when $\mu>0$.
     We use $\mu$ as the energy unit and set $\Lambda = 10^5$.
     The black solid line is the result of numerical calculation of $E^{\text{MF}}$ according to Eqs.~\eqref{eq:tree} and \eqref{eq:4}.
     The red dashed line represents Eq.~\eqref{eq:12} with $u=v=\Delta/2$.
     The general behavior of the effective potential over a much wider range of $\Delta$ (up to $\Delta = 50 \sqrt{\mu}$) is shown in the inset.
     One can see that Eq.~\eqref{eq:12} is a good approximation of the mean-field effective potential.
   The dotted line marks the position of the mean-field stable solution around $\Delta = \sqrt{\mu}$, which is the global minimum.}
 \end{figure}

For a 2D $p$-wave resonant Fermi gas in the strong-coupling limit where $g^2 \ln \frac{\Lambda}{k_F} \gg 1$,
the leading order of $E^{\text{MF}}$ has a logarithmic dependence on $\Lambda$,
which can be cast in the following form,
\begin{align}
\label{eq:6}
 E^{\text{MF}} \approx  \frac{1}{2\pi}\! \ln \frac{\Lambda}{f_1(\boldsymbol{\Delta},\mu)} \left[ {2 (|\Delta|^2)^2 + \left| \boldsymbol{\Delta} \cdot \boldsymbol{\Delta} \right|^2}
                         - {2 \mu |\Delta|^2} \right],
\end{align}
where
\begin{align}
  \label{eq:f1}
  f_1(\boldsymbol{\Delta},\mu) = & \sqrt{\frac{|\Delta|^2 + |\boldsymbol{\Delta} \cdot \boldsymbol{\Delta}|- \theta(-\mu) \mu}{2}}  \nonumber \\
&  + \sqrt{\frac{|\Delta|^2 - |\boldsymbol{\Delta} \cdot \boldsymbol{\Delta}| - \theta(-\mu) \mu}{2}}.
\end{align}
To derive the above Eq.~\eqref{eq:6}, we have taken into account the resonance condition $\delta_R = 0$.
Eq.~\eqref{eq:6} is manifestly invariant under $U(1)$ gauge transformation and $SO(2)$ rotation in the $xy$-plane, which is consistent with the symmetry of the Hamiltonian (Eq.~\eqref{eq:2}).
Up to a gauge transformation, the complex vector $\boldsymbol{\Delta}$ can always be written as $\boldsymbol{\Delta} = \boldsymbol{u} + i \boldsymbol{v}$, where $\boldsymbol{u}$ and $\boldsymbol{v}$ are real vectors satisfying $\boldsymbol{u} \cdot \boldsymbol{v} =0$.
Furthermore, we can isolate the most divergent contribution in Eq.~\eqref{eq:6}, which is proportional to $\ln \Lambda$, by replacing $f_1$ with $\sqrt{|\mu|}$.
This extracts the most important dependence of $E^{\text{MF}}$ on $\boldsymbol{\Delta}$ but only introduces negligible difference proportional to $\ln^0 \Lambda$ (see Fig.~\ref{fig:mf}).
The effective potential can then be written as
\begin{equation}
  \label{eq:12}
E^{\text{MF}} = \frac{1}{2\pi} \ln \frac{\Lambda}{\sqrt{|\mu|}} [3u^4 +3 v^4 + 2u^2 v^2 - 2 \mu (u^2 + v^2)].
\end{equation}
It is straightforward to show that the effective potential has a local minimum in the $u-v$ space at $u^2 = v^2 = \mu /4 $ (which is also the global minimum) when $\mu>0$ (see Fig.~\ref{fig:mf}).
Without loss of generality, we can choose the basis $\{ \boldsymbol{e}_x, \boldsymbol{e}_y \}$ so that
\begin{equation}
  \label{eq:mfd}
  \boldsymbol{\Delta}^{\text{MF}} = \frac{\sqrt{ \mu}}{2} (\boldsymbol{e}_x \pm i \boldsymbol{e}_y).
  \end{equation}
The pairing gap is $2|\boldsymbol{\Delta} \cdot \boldsymbol{k}| = \sqrt{\mu} |\boldsymbol{k}|$.
Therefore, at mean-field level, we expect a fully gapped $p \pm ip$ superfluid for a 2D Fermi gas near a strong-coupling $p$-wave resonance.
The mean-field equation of state can further be obtained by considering the number equation $n = - \partial E/ \partial \mu$ at the minimum obtained above, which in the leading order yields $\mu^{\text{MF}} \approx \frac{k_{F}^2}{2 \ln \Lambda/k_F} $~\cite{Pfaffian}.
The compressibility is thus
\begin{equation}
  \kappa^{\text{MF}} = \left( n^2 \frac{\partial \mu}{\partial n}\right)^{-1} \approx \frac{\ln \Lambda/k_F}{2\pi n^2},
  \end{equation}
  which is positive, indicating a stable superfluid.

 \begin{figure}
   \includegraphics[width=3.5in]{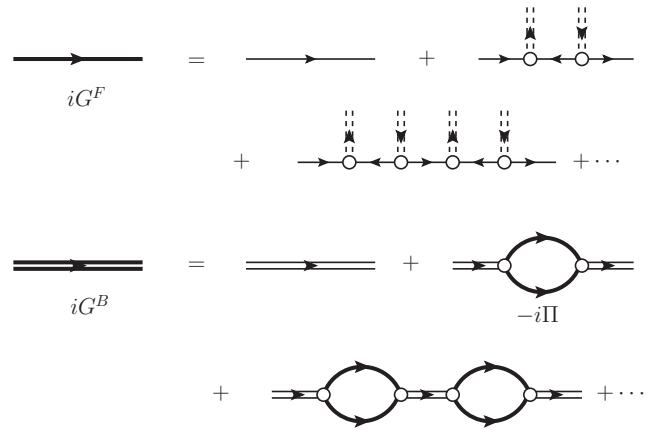}
   \caption{\label{fig:boson}The fermion propagator in the presence of pairing and the dressed boson propagator. The former is represented by a thick solid line, and the (dressed) boson propagator is represented by a (thick) double solid line (see also the caption of Fig.~\ref{fig:1loop}).}
 \end{figure}

\begin{figure}
  \includegraphics[width=3in]{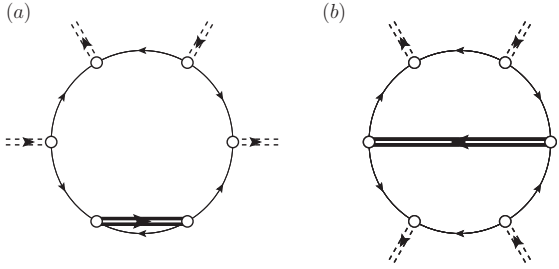}
  \caption{\label{fig:2loop}Two-loop diagrams with $N_{\Delta} = 4$ (four pairing fields represented by double dashed lines) in terms of dressed boson propagators (see also the captions of Figs.~\ref{fig:1loop} and \ref{fig:boson}). Diagram (a) contributes at the order of $\ln^{-1} \Lambda$, while (b) contributes at the order of $\ln \Lambda$, which is the most divergent two-loop diagram. }
\end{figure}

\section{Beyond-mean-field contributions}
We now discuss the effect of the quantum fluctuations on the mean-field $p \pm ip$ pairing, and will choose $\boldsymbol{\Delta} = \Delta/2(\boldsymbol{e}_x + i \boldsymbol{e}_y)$.
Diagrammatically, beyond-mean-field contributions of quantum fluctuations to the effective potential are given by diagrams with two or more loops.
For these diagrams, the bosonic propagator is an important building block.
It is defined as
\begin{equation}
G^B_{ij}(Q) \equiv -i \int_{-\infty}^{\infty} dt e^{iQ_0 t} \langle T b_{i,\boldsymbol{Q}}(t) b^{\dagger}_{j,\boldsymbol{Q}}(0) \rangle ,
\end{equation}
where $Q = (Q_0, \boldsymbol{Q})$ represents the frequency and momentum, $b$ and $b^{\dagger}$ are time-dependent Heisenberg-picture operators, $T$ enforces time ordering, and the expectation value is calculated with respect to the ground state.
In the absence of interchannel coupling, the bare bosonic propagator is
\begin{equation}
G^{B(0)} = \delta_{ij} (Q_0 - \boldsymbol{Q}^{2}/4 + 2 \mu + i 0^+)^{-1}.
\end{equation}
When the coupling is strong, it is heavily dressed due to the strong hybridization of the closed and open channels.
The dressing effect can be summarized using a self energy of the closed-channel bosons,
and the dominant contribution to the self energy, $\Pi$, is shown in Fig.~\ref{fig:boson}, which contains a UV logarithmic divergence, i.e., $\Pi \sim \ln \Lambda$.
The dressed boson propagator is then obtained via
\begin{equation}
G^B = [(G^{B(0)})^{-1} - \Pi]^{-1} .
\end{equation}
In the strong-coupling limit when $g^2 \ln\frac{\Lambda}{k_F}\gg1 $, the leading order of $G^B$ can be calculated as,
\begin{align}
  \label{eq:7}
  G^{B}_{ij}(Q)  \approx & -4 \pi g^{-2} \delta_{ij} \left[ \left(\frac{\boldsymbol{Q}^2}{4} -Q_0 - 2 \mu + 4\Delta^2 - i 0^+ \right) \right. \nonumber \\
                   & \left. \times \ln \frac{\Lambda^2}{h(Q_0+i0^+,\boldsymbol{Q}^2,\mu,\Delta)} \right]^{-1},
\end{align}
where $h$ is a regular function with the dimension of energy and does not have any poles on the real axis or the upper half plane in the complex $Q_0$-plane.
When $Q_0$ and $\boldsymbol{Q}^2$ are much larger than $\mu$ and $\Delta^2$, $h\approx -Q_0 + \boldsymbol{Q}^2/4 - i0^+$.
Compared with the bare boson propagator, the dressed propagator $G^B$ (Eq.~\eqref{eq:7}) behaves similarly but with a slow logarithmic dependence in the denominator and a $g^{-2}$ factor.
For fermions, $G^B$ mediates an effective $p$-wave interaction with a dispersive coupling, $g^2 G^B$; the logarithm in the denominator is a characteristic behavior of the $p$-wave scattering $T$-matrix (see Eq.~\eqref{eq:1}).
This suggests that the effective interaction between fermions in the strong interchannel coupling limit is logarithmically small.
It appears in all diagrams beyond the mean field or one-loop diagrams.
Superficially, one might then argue that such a small coupling suppresses quantum fluctuations from higher order processes.
However, such suppression only exists if loop integrals are all regular, i.e., do not have further UV divergences or logarithmic divergences.
If the loop integrals further contain logarithmic divergences which might offset the logarithmic suppression in $G^B$, the quantum fluctuations from these loops can lead to a net contribution comparable to mean-field energy.

Contributions beyond mean field starts with two-loop diagrams with four $\boldsymbol{\Delta}$ fields.
The two diagrams of this kind are shown in Fig.~\ref{fig:2loop}.
A direct power counting suggests that Fig.~\ref{fig:2loop}(a) $\sim \ln^{-1} \Lambda$ while Fig.~\ref{fig:2loop}(b) $\sim \ln \Lambda$ which is of the same order of $E^{\text{MF}}$.
This is exactly due to divergent loop integrals in Fig.~\ref{fig:2loop}(b) as we discussed in the previous paragraph.
More specifically, each of the two loop integrals in Fig.~\ref{fig:2loop}(b) is logarithmically divergent when the momentum of the other loop is fixed, which yields $\ln^2 \Lambda$.
This leads to an overall scaling behavior of $\ln \Lambda$ for this diagram when the $\ln^{-1} \Lambda$ factor in the boson propagator is taken into account.
Using Eq.~\eqref{eq:7}, one can calculate this two-loop contribution to the leading order of $1/\ln \frac{\Lambda}{k_F}$, which yields
\begin{equation}
  \label{eq:e2}
  E^{(2)} \approx - \frac{2}{3 \pi} \Delta^4 \ln \frac{\Lambda}{f_2({\Delta},\mu)} \ln \ln \frac{\Lambda}{f_3({\Delta},\mu)},
\end{equation}
where we have regulated the infrared contributions near Fermi surface by introducing a $f_2$ function.
Both $f_{2}$ and $f_3$ are regular functions with dimension of momentum; $f_2$ scales as $\Delta$ when $\mu>0$ and as max\{$\Delta,\sqrt{|\mu|}$\} when $\mu<0$ (see below for more discussions).
This is consistent with our scaling analysis except for the additional yet much slower factor $\ln \ln \Lambda$ which comes from the interplay between the loop integral and the logarithmic factor in the boson propagator (see Eq.~\eqref{eq:7}).

 \begin{figure}
   \includegraphics[width=3.5in]{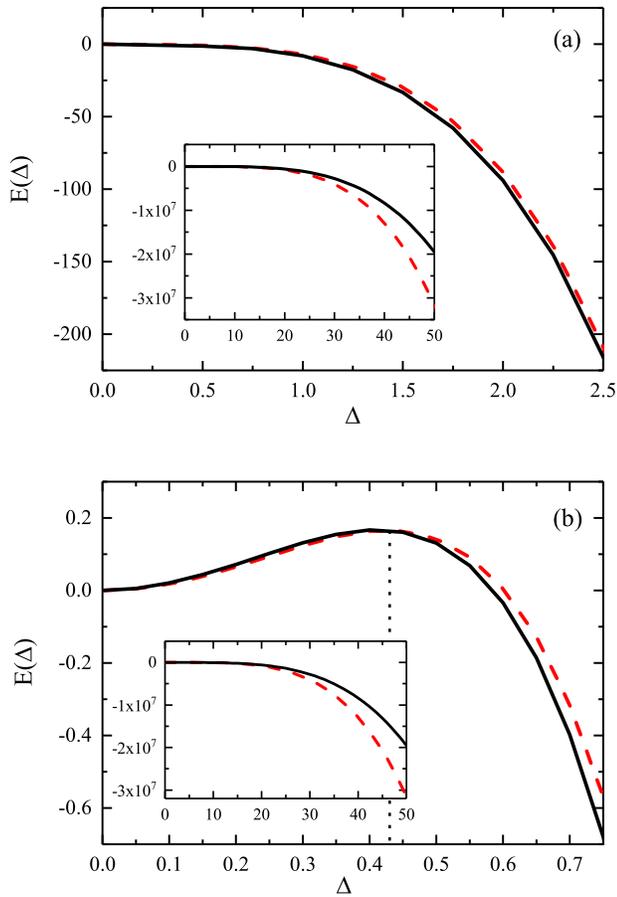}
   \caption{\label{fig:with2loop}The effective potential in the two-loop order when (a) $\mu>0$ and (b) $\mu<0$.
     We chose similar parameters as Fig.~\ref{fig:mf}, except that the energy unit is now $|\mu|$.
     The black solid lines show the numerical result of $E^{\text{MF}} + E^{(2)}$.
     The red dashed lines are obtained by setting $f_{1,2,3} = \sqrt{|\mu|}$;
     This is effectively to isolated the most divergent contributions in Eq.~\eqref{eq:13}, which is proportional to $\ln \Lambda$,  and neglect the subleading contributions which does not depend on $\ln \Lambda$.
     The insets show the general behavior of the effective potential over a much wider range of $\Delta$.
     One can see that the choice of $f_{1,2,3} = \sqrt{|\mu|}$ does not change the qualitative picture.
   The dotted line marks the position of the unstable solution when $\mu<0$ which is a local maximum.}
 \end{figure}

It can be seen that $E^{(2)}$ overtakes $E^{\text{MF}}$ (Eq.~\eqref{eq:6}) parametrically by the factor of $\ln \ln \frac{\Lambda}{f_3}$.
Furthermore, $E^{(2)}$ is negative because the integrals have a large contribution from the UV regime, where the effective interaction between fermions ($\sim g^2 (\boldsymbol{p}-\boldsymbol{q})^2 G^B $) is attractive.
We now obtain the leading order of the effective potential in the strong-coupling limit:
\begin{align}
  \label{eq:13}
  E & =  E^{\text{MF}} + E^{(2)} \nonumber \\
    & \approx \frac{1}{2\pi} \ln \frac{\Lambda}{f_1(\boldsymbol{\Delta},\mu)} \left[ - \left( \frac{4}{3} \ln \ln \frac{\Lambda}{f_3} - \frac{1}{2}  \right) \Delta^4 - \mu \Delta^2 \right] \nonumber \\
  & + \frac{2}{3 \pi} \Delta^4 \ln \frac{f_2}{f_1} \ln \ln \frac{\Lambda}{f_3}.
\end{align}
Following Eq.~\eqref{eq:f1} and the discussions after Eq.~\eqref{eq:e2}, one anticipates that $f_2/f_1$ is of order of unity, so the last term in Eq.~\eqref{eq:13} is of order of $\ln^0 \Lambda$ and can be neglected compared with other terms.
Since $\Lambda \gg \Delta, |\mu|^{1/2}$, the coefficient of the quartic term in the effective potential is negative due to the $\ln \ln \Lambda/f_3$ factor, in contract to a positive one in the standard Ginzburg-Landau free energy.
This implies that the mean-field pairing superfluid is destabilized by quantum fluctuations.
We plot the effective potential including the two-loop contribution in Fig.~\ref{fig:with2loop}.
One can see that there is no non-trivial solution when $\mu>0$.
When $\mu<0$, the effective potential has a local maximum, which corresponds to an unstable solution.
As this instability is driven by the large $\ln \Lambda$ and $\ln \ln \Lambda$ factors in the two-loop contribution, our conclusion on the instability is insensitive to the details of $f_{2,3}$.
In other words, since the contribution of quantum fluctuations is mainly from a low-energy many-body momentum scale all the way to the UV scale $\Lambda$, different choices of the infrared cutoff scale only yield minor quantitative difference in the final result as $\Lambda$ is much larger than either $\Delta$ or $\sqrt{|\mu|}$.
In fact, the difference is subleading, of relative order of $\ln^{-1}\Lambda$.
This has been verified in the numerical calculations presented in Fig.~\ref{fig:with2loop}.
In the leading order of $\ln^{-1}\Lambda$, the equation of state at the stationary solution when $\mu<0$ (which is a local maximum) is $\mu = - \frac{2}{3} \frac{\ln \ln \Lambda/k_F}{\ln \Lambda/k_F} k_F^2$.
One can further calculate the compressibility at the local maximum to obtain:
\begin{align}
  \label{eq:5}
  \kappa^{\text{2-loop}} = - \frac{3}{16\pi} \frac{\ln  \Lambda/k_F}{\ln \ln \Lambda/k_F} n^{-2}, 
\end{align}
which is negative, indeed showing a thermodynamical instability.

For comparison, we apply the same approach to superfluids at a $p$-wave resonance in the weak interchannel coupling limit where $g^2 \ln \frac{\Lambda}{k_F} \ll 1$ (but still with $\delta_R =0$).
In this limit, the fluctuations are perturbative in terms of $g^2$.
Therefore, the mean field gives the leading order contribution to the effective potential, which can be calculated as,
\begin{equation}
  \label{eq:14}
  E \approx \frac{1}{2\pi} \ln \frac{\Lambda}{f_1(\boldsymbol{\Delta},\mu)} \left[ {2 (|\Delta|^2)^2 + \left| \boldsymbol{\Delta} \cdot \boldsymbol{\Delta} \right|^2} \right] - \frac{2 \mu \left| \Delta \right|^2}{g^2}
\end{equation}
Different from the strong-coupling limit (Eq.~\eqref{eq:13}), it has a local minimum when $\mu>0$.
In the leading order, the solution can be calculated as,
\begin{align}
  \mu & = \frac{1}{2 \pi} \ln \frac{\Lambda}{k_F}  {g^4} n, \nonumber \\
  \boldsymbol{\Delta} & = \frac{g \sqrt{n}}{2}(\boldsymbol{e}_x \pm \boldsymbol{e}_y),
\end{align}
and the compressibility is 
\begin{equation}
  \label{eq:10}
  \kappa^{\text{weak}} = \frac{2 \pi}{g^4 \ln \Lambda/k_F} n^{-2}.
\end{equation}
This corresponds to a stable $p+ip$ superfluid at resonance in the weak interchannel coupling limit, which is consistent with previous studies~\cite{Gurarie2007}.
Therefore, at resonance when the interchannel coupling is increased, one should expect an onset of instability in the $p$-wave superfluid near $g^2 \sim \ln^{-1} \frac{\Lambda}{k_F}$.

\begin{figure}
  \includegraphics[width=3in]{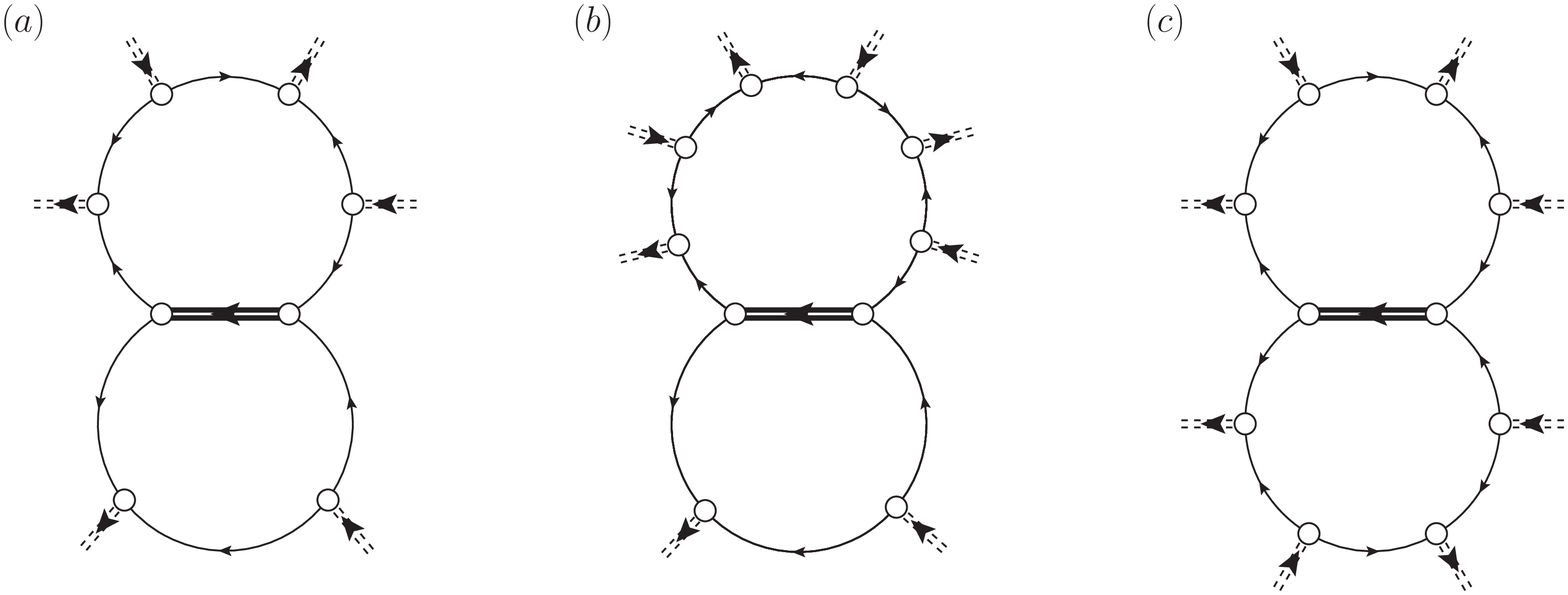}
  \caption{\label{fig:2loop-m}Examples of two-loop diagrams with $N_{\Delta} = 6$ (diagram (a)) and $N_{\Delta} = 8$ (diagrams (b) and (c)) (see also the captions of Figs.~\ref{fig:1loop} and \ref{fig:boson}). They contribute to the effective potential at the order of $\ln^0 \Lambda$, $\ln^0 \Lambda$, and $\ln ^{-1} \Lambda$, respectively, which are in higher order compared with Fig.~\ref{fig:2loop}(b).}
\end{figure}

Before concluding this part of discussion, let us make a remark about other two-loop contributions.
All two-loop diagrams can be classified in terms of the number of pairing fields, $N_{\Delta}$, which is an even integer.
The $N_{\Delta} = 2$ case is a reducible diagram and thus does not contribute to the effective potential.
The above discussion in this section has been focusing on $N_{\Delta} = 4$.
For two-loop diagrams with $N_{\Delta} \ge 6$, they do not contain overall UV divergences and therefore smaller than the mean-field result and Fig.~\ref{fig:2loop}(b) by a factor of $\ln^{-\alpha}\Lambda$ ($\alpha \ge 1$ and can depend on the topology of the diagram as shown in Fig.~\ref{fig:2loop-m}).

One might further argue that terms of the form $\Delta^{6,8,\cdots}$~\cite{gauge} in the effective potential with power of $\Delta$ higher than 4, although smaller in terms of $\ln ^{-1} \Lambda$, might lead to extra stationary solutions where $\Delta^2/\mu \gg 1$.
Note that the effective potential as a function of $\Delta$, after proper resummation of diagrams, shall have a highly restricted form, because of the general scaling relation between $\Delta^4$ and the effective potential per unit volume in 2D.
Although diagrams with $N_{\Delta} = 6,8,\cdots$ (see Fig.~\ref{fig:2loop-m} for some examples) superficially imply terms of order of $\Delta^{6,8,\cdots}$ in the effective potential, the contributions from these diagrams are dominated by infrared fluctuations and can only be regularized by $\Delta$ in the limit if $\Delta^2 \gg \mu$.
As a result of the infrared regularization, these diagrams do not yield terms of the form $\Delta^{6,8,\cdots}$.
This can be easily visualized in the limit when $\Delta^2 \gg \mu$ or $\mu=0$.
In this limit, $\Delta$ is the only relevant momentum scale in all loop integrals.
For the strong coupling resonance we are interested in, a simple dimension analysis leads to a contribution proportional to $\Delta^4$.
In other words, when $\Delta^2 \gg \mu$, the highest power of $\Delta$ in the effective potential is four.
Summation of diagrams with $N_{\Delta} \ge 6$ will lead to a net contribution scaling as $\Delta^4$, but of the order of $\ln ^0 \Lambda$.
Compared with the leading term discussed above which is of the order of $\ln \Lambda$, these contributions effectively slightly modify the infrared cutoff of the logarithm, i.e., $f_2$ which we have employed before.
Furthermore, according to previous paragraph, this modification is in higher order of $\ln^{-1} \Lambda$ compared with Fig.~\ref{fig:2loop}(b), thus all irrelevant for our discussions of the instability.

\section{Discussions}
At a $p$-wave resonance in the limit of strong interchannel coupling under consideration in the article, the two-body scattering phase shift is logarithmically small (see Eq.~\eqref{eq:1}), implying weak effective interchannel coupling at low energy scale.
One might speculate that the mean-field theory shall be adequate for such a weakly interacting system.
This is in fact incorrect as the small phase shift or weak effective interchannel coupling is a necessary condition for the mean-field theory to be correct but not a sufficient condition.

From the loop-expansion point of view, the mean-field theory or the one-loop theory is only valid when higher-loop contributions are parametrically small.
This requires:

1) effective interaction ($\sim g^2 G^B$, see Eq.~\eqref{eq:7} and Fig.~\ref{fig:boson}) is small or equivalently the phase shift is logarithmically small in our case (see Eq.~\eqref{eq:1});

2) all diagrams with two or more loops do not have other logarithmic divergences so as not to offset the smallness of the effective interaction~\cite{marginal}.

This is indeed exactly the case for pairing states near a 4D $s$-wave resonance studied before~\cite{Nishida2006}.
However, it is not the case for a 2D $p$-wave resonance under consideration.
The explicit result on two-loop fluctuations obtained in this article demonstrates that condition 2) has been violated and the mean-field pairing state is very fragile and vulnerable to quantum fluctuations.
Considering the weak-coupling limit studied before, where the fluctuation effect is perturbative and a $p+ip$ superfluid is stable, one can expect an onset of instability near $g^2 \sim \ln^{-1} \frac{\Lambda}{k_F}$ when going from weak- to strong-resonance limit.
Due to the competition between $g^2$ and $\ln^{-1} \frac{\Lambda}{k_F}$, the instability can also be driven by lowering the particle density with a fixed interchannel coupling.

In the limit $g \rightarrow \infty$, the two-channel model is reduced to a one-channel model with tunable attractive interparticle interactions~\cite{onechannel}.
Our analysis thus suggests that, for a one-channel model, the same instability exists if the two-body scattering is tuned to resonance.
In a quasi-2D system where the  motion of the particles in $z$ direction is tightly confined while the motion in $x,y$ directions is extended, the onset of instability occurs when $g^2 \sim \ln^{-1}\frac{1}{k_Fa_{0}}$ if $\Lambda/k_F \gg 1$ and $\Lambda a_0 \sim 1$ ($a_0$ is the confinement radius in $z$ direction)~\cite{quasi2d}.

Beyond the two-loop calculation presented, one can further identify a family of diagrams containing more loops that are also marginal and represent other potential logarithmic contributions from quantum fluctuations.
A resummation is needed to further address the issue of stability beyond what is presented here for $L=2$ contributions, with all marginal contributions fully taken into account.
Quantitative calculations along this direction remain to be carried out in the future.
Nevertheless, we speculate that the mean-field pairing solution in general is unstable against these additional fluctuations and a many-body pairing ground state, if it exists, shall only be found with $\Delta$ being much larger than the Fermi momentum in the problem. 
This makes 2D $p$-wave superfluids near resonance a unique class of quantum many-body states, highly non-mean-field in nature or non-BCS like.

\begin{acknowledgments}
  We thank Jeff Maki for helping improve the manuscript.
This work is in part supported by the Canadian Institute for 
Advanced Research and the Natural Sciences and Engineering 
Research Council of Canada (a Discovery Grant).
\end{acknowledgments}


\begin{thebibliography}{References}
\bibitem{weight}From the closed-channel point of view, Eq.~\eqref{eq:1} indicates that the propagator of a closed-channel molecule with zero momentum for a two-body problem has the form $G^B(E) \sim [g^2/s + E + g^2/(4\pi) E \ln (\Lambda^2/(-E))]^{-1}$, where $s$ is the scattering area determined through $s^{-1} = -\delta/g^2 + \Lambda^2/(4 \pi)$. $G^B$ has a pole on the negative real axis of the complex $E-$plane, which corresponds to the energy of a bound state. One can further calculate the residue of $G^B$ at this pole to get $Z \sim [g^2 \ln (\Lambda^2s)]^{-1}$. This shows that the weight in the closed channel is negligible in the strong coupling limit and goes to zero at resonance when $s=\infty$.
\bibitem{gauge} Here we consider $\boldsymbol{\Delta}$ of the form $\Delta/2(\boldsymbol{e}_x + i \boldsymbol{e}_y)$. For a general complex vector $\boldsymbol{\Delta}$, these terms shall be combinations of gauge invariant forms of $\boldsymbol{\Delta}$, such as $|\Delta|^2 \equiv \boldsymbol{\Delta} \cdot \boldsymbol{\Delta}^{*}$ and $|\boldsymbol{\Delta} \cdot \boldsymbol{\Delta}|^2$, with the power $6,8,\cdots$ representing the power of $\boldsymbol{\Delta}$.
\bibitem{marginal}This requirement is equivalent to the statement that there are no other marginal operators in the theory apart from the two-body operator discussed in requirement 1).
    Alternatively, the sufficient condition for the mean-field theory to be valid is that the only marginal operator in the theory is the two-body one.
\bibitem{onechannel}When $g \rightarrow \infty$,  Eq.~\eqref{eq:1} reduces to the standard $p$-wave scattering $T$-matrix in 2D, $T = \boldsymbol{k} \cdot \boldsymbol{k}^{\prime} \left[ s^{-1} + E/(4\pi) \ln (\Lambda^2/E) + i E/4 \right]^{-1}$, where $s$ is the scattering area. The resonance corresponds to $s = \infty$.
\end{thebibliography}
\end{document}